# Molecular Beam Epitaxial Growth of Bi$_2$Te$_3$ and Sb$_2$Te$_3$ Topological Insulators on GaAs (111) Substrates: A Potential Route to Fabricate Topological Insulator p-n Junction

Zhaoquan Zeng[1,†], Timothy A. Morgan[1], Dongsheng Fan[1,2], Chen Li[1], Yusuke Hirono[1], Xian Hu[1], Yanfei Zhao[3], Joon Sue Lee[4], Jian Wang[3,4*], Zhiming M. Wang[1,5,6*], Shuiqing Yu[1,2], Michael E. Hawkridge[1], Mourad Benamara[1], and Gregory J. Salamo[1]

[1] Arkansas Institute for Nanoscale Material Sciences and Engineering, University of Arkansas, Fayetteville, AR 72701, USA
[2] Department of Electrical Engineering, University of Arkansas, Fayetteville, AR 72701, USA
[3] International Center for Quantum Materials, School of Physics, Peking University, Beijing, 100871, China
[4] The Center for Nanoscale Science and Department of Physics, The Pennsylvania State University, University Park, PA 16802, USA
[5] State Key Laboratory of Electronic Thin Films and Integrated Devices, University of Electronic Science and Technology of China, Chengdu 610054, China
[6] Engineering Research Center for Semiconductor Integrated Technology, Institute of Semiconductors, Chinese Academy of Science Beijing 100083, China
[†] Present address: Electrical and Computer Engineering Department, The Ohio State University, Columbus, Ohio 43210, USA

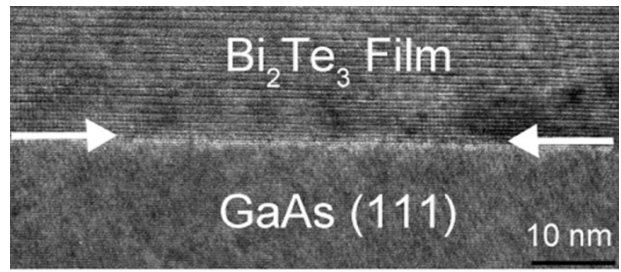

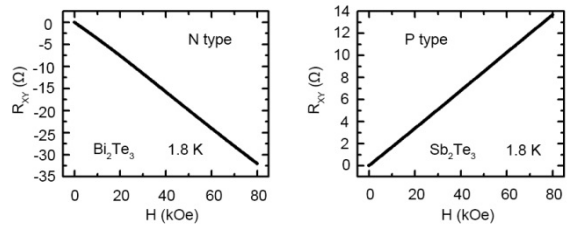

Molecular beam epitaxial growth of high quality Q$_2$Te$_3$ topological insulator (TI) films was systematically investigated on both vicinal and non-vicinal GaAs (111) substrate. Transport measurements indicate that these as-grown Q$_2$Te$_3$ films may be used for the fabrication of topological insulator p-n junctions, making it possible to integrate TI-based devices with III-V semiconductor devices.

Provide the authors' website if possible.
Jian Wang, http://www.phy.pku.edu.cn/icqmjianwanggroup/people/people-wj.html
Gregory J. Salamo, http://www.uark.edu/misc/salamo/

# Molecular Beam Epitaxial Growth of Bi$_2$Te$_3$ and Sb$_2$Te$_3$ Topological Insulators on GaAs (111) Substrates: A Potential Route to Fabricate Topological Insulator p-n Junction

Zhaoquan Zeng[1,†], Timothy A. Morgan[1], Dongsheng Fan[1,2], Chen Li[1], Yusuke Hirono[1], Xian Hu[1], Yanfei Zhao[3], Joon Sue Lee[4], Jian Wang[3,4](✉), Zhiming M. Wang[1,5,6](✉), Shuiqing Yu[1,2], Michael E. Hawkridge[1], Mourad Benamara[1], and Gregory J. Salamo[1]

[1] Arkansas Institute for Nanoscale Material Sciences and Engineering, University of Arkansas, Fayetteville, AR 72701, USA

[2] Department of Electrical Engineering, University of Arkansas, Fayetteville, AR 72701, USA

[3] International Center for Quantum Materials, School of Physics, Peking University, Beijing, 100871, China

[4] The Center for Nanoscale Science and Department of Physics, The Pennsylvania State University, University Park, PA 16802, USA

[5] State Key Laboratory of Electronic Thin Films and Integrated Devices, University of Electronic Science and Technology of China, Chengdu 610054, China

[6] Engineering Research Center for Semiconductor Integrated Technology, Institute of Semiconductors, Chinese Academy of Science Beijing 100083, China

[†] Present address: Electrical and Computer Engineering Department, The Ohio State University, Columbus, Ohio 43210, USA

## ABSTRACT

High quality Bi$_2$Te$_3$ and Sb$_2$Te$_3$ topological insulators films were epitaxially grown on GaAs (111) substrate using solid source molecular beam epitaxy. Their growth and behavior on both vicinal and non-vicinal GaAs (111) substrates were investigated by reflection high-energy electron diffraction, atomic force microscopy, X-ray diffraction, and high resolution transmission electron microscopy. It is found that non-vicinal GaAs (111) substrate is better than a vicinal substrate to provide high quality Bi$_2$Te$_3$ and Sb$_2$Te$_3$ films. Hall and magnetoresistance measurements indicate that p type Sb$_2$Te$_3$ and n type Bi$_2$Te$_3$ topological insulator films can be directly grown on a GaAs (111) substrate, which may pave a way to fabricate topological insulator p-n junction on the same substrate, compatible with the fabrication process of present semiconductor optoelectronic devices.

---

Address correspondence to Jian Wang, email: jianwangphysics@pku.edu.cn; Zhiming M. Wang, email: zhmwang@semi.ac.cn

## 1. Introduction

$Q_2Te_3$ (Q = Bi, Sb) is a typical V-VI narrow gap semiconductor, having a rhombohedral unit cell that can be considered as groupings of -Te-Q-Te-Q-Te- planes, referred to as quintuple layers (QLs). Within the QL unit, the chemical bond provides the binding force, while the weak van der Waals (vdW) force acts between adjacent QLs. $Q_2Te_3$ is a traditional thermoelectric (TE) material (e.g. ZT ≈ 1 for $Bi_2Te_3$ at 300 K), [1] that has caused attention in the past decade due to its superior TE performance. Recently, $Q_2Te_3$-based research has increased because of the discovery of a new state of matter known as a three-dimensional (3D) "topological insulator" (TI). This material is insulating in bulk with a finite band gap but possesses a gapless surface state protected by time reversal symmetry (TRS), [2] which has been confirmed by angle-resolved photoemission spectroscopy (ARPES) [3-5] and transport measurements. [6-10] The emergence of TIs has brought an increase in the study of exotic quantum physics, such as Majorana fermions and magnetic monopoles, [11-13] which may pave the way for quantum computation applications. These investigations are also accelerating the development of spintronics due to the spin helical structure of the TI surface state. [14] Additionally, a very recent prediction shows that TI may be a promising candidate for use in high performance photodetectors in the terahertz (THz) to infrared (IR) frequency range because of high absorbance. [15]

With so many potential applications, it has become important to understand how to fabricate these materials. For example, molecular beam epitaxy (MBE) is one way to grow high quality TIs due to the advantages of ultra-high vacuum (UHV) background pressure and precise control of growth parameters. [5] In this work, we have systematically investigated the fabrication of high-quality $Q_2Te_3$ films by MBE on GaAs substrates. Up to now, Si, $SrTiO_3$, sapphire, and graphene have already been used as substrates for the growth of $Q_2Te_3$ films by MBE. [3, 16, 17, 18] There have also been a few investigations on the MBE growth of $Q_2Te_3$ on a GaAs substrate. For example, Liu et al. demonstrated epitaxial growth of both $Bi_2Te_3$ and $Bi_2Se_3$ films on GaAs (100) directly by van der Waals epitaxy (vdWe). [19] They reported that rotation domains formed in the TIs films due to the lattice symmetry mismatch between the hexagonal lattices of $Bi_2Te_3$ and the cubic symmetry of the GaAs (001) surface. [19] The vdWe growth method has also been used for the growth of $Bi_2Se_3$ on a vicinal Si (111) surface. [20] However, an amorphous intermediate layer formed in this approach which of course is not good for the coupling of spins between the substrate and the TI film, and could hinder the integration of TIs on present semiconductor optoelectronic devices. On the other hand, GaAs (111) should be a good substrate to realize coherent heteroepitaxy of $Q_2Te_3$ due to the lattice symmetry. For example, He et al. focused on the measurement of the magnetoresistance properties of $Bi_2Te_3$ thin films grown on semi-insulating GaAs (111) with ZnSe as a buffer layer [21] while Richardella et al. demonstrated coherent heteroepitaxy of $Bi_2Se_3$ on GaAs (111)B. [22]

In this paper, we investigate the growth and behavior of $Q_2Te_3$ on both vicinal and non-vicinal GaAs (111) (V-GaAs (111) and Nv-GaAs(111)) substrates by reflection high-energy electron diffraction (RHEED), atomic force microscopy (AFM), X-ray diffraction (XRD) and high resolution transmission electron microscopy (HRTEM). Together with Hall and magnetoresistance (MR) measurements, our observations indicate that p type $Sb_2Te_3$ and n type $Bi_2Te_3$ topological insulator films can be directly grown on a GaAs (111) substrate, enhancing the opportunity to fabricate topological insulator p-n junctions. [23]

## 2. Experimental

A Riber 32P solid source MBE system was used to grow $Q_2Te_3$ films on epi-ready GaAs (111) substrates. The growth chamber is equipped with a RHEED (Staib) and a transmission optical system (kSA BandiT) monitoring the substrate band edge to give accurate growth temperature down to 150 °C. High purity 6N Bi, Sb and Te sources were evaporated to provide the beam fluxes for the film growth with a ratio of about 1:12. The morphology of as-grown films were quenched by turning off the manipulator heater right after the growth, and a Veeco ambient AFM was used to characterize the sample surface. High resolution XRD (HRXRD) was performed on a PANalytical X'Pert MRD system equipped with a parabolic mirror and PIXcel™ detector. The interface was characterized on FEI Titan 80-300 TEM. The Hall bar structure for transport measurement was fabricated using standard photo-lithography method (see Supporting Information for more details).

The $Q_2Te_3$ films were grown on GaAs (111)A and GaAs (111)B substrates. As a result, two kinds of different RHEED evolutions were observed during growth due to the difference between A and B planes of GaAs (111) (see Supporting Information for more details). As a comparison, both V-GaAs (111) and Nv-GaAs(111) substrates were used as substrates for growth of $Q_2Te_3$ films. This is motivated by the fact that a vicinal Si (111) substrate was observed to improve the quality of $Bi_2Se_3$ epitaxial films grown by vdWe. [20] After deoxidizing GaAs substrates at 610 °C, a two steps growth method, i.e. a high temperature GaAs buffer layer growth at 590 °C and a low temperature $Q_2Te_3$ film growth at 250 °C, was performed. More than ten kinds of different samples were investigated, and here we will show the data from the six kinds of samples as shown in table 1. Sample A (200 nm $Bi_2Te_3$/V-GaAs (111)A-3°) and sample B (400 nm $Sb_2Te_3$/V-GaAs (111)A-3°) were a 200 nm thick $Bi_2Te_3$ film and a 400 nm thick $Sb_2Te_3$ film grown on V-GaAs (111)A substrate (3° miscut) with a 330 nm thick GaAs buffer layer. Samples C and D were 30 nm thick $Bi_2Te_3$ films grown on V-GaAs (111)A and Nv-GaAs (111)A with the same thick GaAs buffer layer (3 nm). Samples E (200 nm thick $Bi_2Te_3$) and F (400 nm thick $Sb_2Te_3$) were grown on V-GaAs (111)B substrate with a 330 nm thick GaAs buffer layer.

## 3. Results and Discussion

### 3.1 Surface morphology studies with atomic force microscopy

Figure 1 shows a set of AFM images, which reveal the different growth modes of $Q_2Te_3$ films observed on both V-GaAs (111) and Nv-GaAs(111) substrates. Figure 1a is the AFM image of sample A (200 nm $Bi_2Te_3$/V-GaAs (111)A-3°). The big terraces and steps run nearly parallel. In most areas, about 1 μm wide terraces and 35 nm high steps can be found. The root-mean-square (RMS) roughness is about 8 nm for 5×5 μm² area. Furthermore, small steps with height of about 1 nm were observed on the flat terrace as seen in Figures 1(b-c) (500×500 nm²), consistent with a single QL thickness for each step. This surface morphology covers the entire sample surface based on more AFM measurements at the different positions on the surface of sample A. Sample B (400 nm $Sb_2Te_3$/V-GaAs (111)A-3°) shows the similar surface morphology as shown in Figure 1d. Obviously, this is a typical step-bunched surface, [24] which forms with the big steps and terraces through a step-flow growth mode. Such surface morphology is very different compared to reported result of spiral growth. [20, 25] In order to avoid concerns over variation in the thickness of GaAs buffer layer or $Q_2Te_3$ film, both samples C (30 nm $Bi_2Te_3$/V-GaAs (111)A-3°) and D (30 nm $Bi_2Te_3$/Nv-GaAs (111)A) were grown on the same molybdenum block simultaneously using the same thick GaAs buffer layer (3 nm) and $Q_2Te_3$ (30 nm) films. As shown in Figure 1e, sample C still shows an initial step-bunched surface. However, spiral growth was observed for sample D as shown in Figures 1(f-h). For the as-grown $Q_2Te_3$ film, AFM investigations indicate that a V-GaAs (111) substrate is conducive to the formation of a step bunched surface through the tendentious step-flow growth induced by the high density of steps providing the high energy barrier to grow over the steps in its surface. [26] Nv-GaAs (111) substrates sustain a smooth surface during spiral growth or partial step-flow growth due to the lower step density associated with non-vicinal substrate surface (see more AFM investigation results in the Supporting Information).

### 3.2 Structural studies with X-ray diffraction and transmission electron microscopy

To determine the crystal quality and epitaxial orientation of as-grown $Q_2Te_3$ films, XRD was performed. Figures 2(a-b) are the 2θ-ω scans within 0-80° range for samples E (200 nm $Bi_2Te_3$/V-GaAs (111)B-3°) and F (400 nm $Sb_2Te_3$/V-GaAs (111)B-3°). Only $Q_2Te_3$ (003) and GaAs (111) families of reflections were observed, indicating the high c-axis orientation of $Q_2Te_3$ films along GaAs [111] direction, i.e. $Q_2Te_3$ (001)//GaAs (111). Compared with the peak position of GaAs substrates, the QL thickness (equals to 1/3 of lattice constant c) was given by $d_{QL}$=1.011 nm for $Bi_2Te_3$ and $d_{QL}$=1.007 nm for $Sb_2Te_3$, respectively, consistent with the RHEED observation results that the $Q_2Te_3$ lattice is fully relaxed. The full-width-half-maximum (FWHM) is 0.09° and 0.014° for $Bi_2Te_3$ (006) and $Sb_2Te_3$ (00,15), respectively, as shown in Figures 2(c-d), indicating high-quality growth. To distinguish the subtle difference of as-grown $Q_2Te_3$ films, reciprocal space mapping (RSM) was performed on samples F (400 nm $Sb_2Te_3$/V-GaAs (111)B-3°), C (30 nm $Bi_2Te_3$/V-GaAs (111)A-3°) and D (30 nm $Bi_2Te_3$/Nv-GaAs (111)A). The RSM results of symmetric planes in Figures 2(e-g) distinctly demonstrated the tilted angle of 0.055°, 0.046° and 0.002° between $Q_2Te_3$ (00,18) and GaAs (222), for samples F, C and D, respectively, indicating higher c-axis epitaxial orientation than vdWe on a vicinal substrate. [21] For these three samples, Sample D grown on non-vicinal substrate has a better c-axis orientation than samples C and F. The RSM result of non-symmetric planes of GaAs (440) and $Bi_2Te_3$ ($\bar{2}2,27$) for sample D in Figure 2h presents the in-plane lattice mismatch between $Bi_2Te_3$ film and GaAs (111) substrate is around 9.3%, which is consistent with the RHEED observation of small critical thickness. Besides, the peak of sample F grown on vicinal substrate with thick GaAs buffer layer is much broader than that of samples C and D, indicating worse crystalline quality. The substrate miscut dependence of the tilted orientation angles for these three samples proves that as-grown $Q_2Te_3$ films are sensitive to the surface lattice orientation of the substrates and V-GaAs (111) substrates can reduce their crystalline quality through the unfavorable step bunching process.

To further clarify the effect of V-GaAs (111) substrates on the crystal quality of as-grown $Q_2Te_3$ film, we investigated the interface between $Q_2Te_3$ films and GaAs buffer layer for these three samples by HRTEM. Figure 3(a) gives the HRTEM images of sample F. It is obvious that the interface is disordered and a lot of defects were formed in both the GaAs buffer layer and $Sb_2Te_3$ film. Furthermore, the fast Fourier transform (FFT) analysis near the interface confirmed the RSM result that there is a small tilted angle of about 0.055° between the $Sb_2Te_3$ film and GaAs (111) substrate. For sample C, a similar result is observed in Figure 3(b), i.e. disordered interface and tilted orientation, except for that fewer defects were formed due to the thinner GaAs buffer layer. In contrast, both a sharp interface and nearly parallel orientation were found for sample D, as shown in Figures 3(c-d). Based on these observations, we can conclude that an Nv-GaAs (111) substrate is better to use than a vicinal substrate in order to fabricate high quality $Q_2Te_3$ films on GaAs (111) substrate.

**3.3 Transport measurements of $Q_2Te_3$ thin films**

In order to demonstrate the topological insulator behavior of as-grown $Q_2Te_3$ films, transport measurements were also performed. For our samples, all the $Sb_2Te_3$ films show p type conducting behavior, whereas, all the $Bi_2Te_3$ films show n type conducting behavior. Figure 4 shows the typical transport measurement results of $Q_2Te_3$ films on GaAs (111)B (see the transport data for $Q_2Te_3$ films on GaAs (111)A in Supporting Information). Based on the linear behavior of Hall resistance vs. magnetic field ($R_{xy}$-H) below 30 kOe, the carrier density and mobility can be estimated as $7.81\times10^{18}$ cm$^{-3}$ and 959 cm$^2$/Vs, $9.13\times10^{18}$ cm$^{-3}$ and 781 cm$^2$/Vs for samples E (200 nm $Bi_2Te_3$/V-GaAs (111)B-3°) and F (400 nm $Sb_2Te_3$/V-GaAs (111)B-3°), respectively, which are typical parameters in TI films. [3, 32] Nonlinear Hall behavior was observed over 30 kOe for sample E, as shown in Figures 4(a), which is likely an indication of more than one transport channel (surface and bulk) in the sample. [28] Magnetoresistance (MR) measurements were performed for three different directional fields up to 80 kOe. In the perpendicular field (H), the R(H)/R(0) increases from 1 to 1.73 and 1.56 for samples E and F at 1.8 K, respectively, over the range of 0 to 80 kOe; linear MR was observed above 20 kOe, which may be attributed to the quantum linear MR (QLMR) of the surface states, [27] as shown in Figures 4(c-d). For the parallel field situation (Figures 4(e-f)), R(H)/R(0) increases to 1.15~1.30 at 80 kOe, which are smaller than that in perpendicular field. In a close-up view as shown in Figure 4g, a clear magnetoconductance peak ($\Delta\sigma$) was observed for sample E in small perpendicular field regime, which can be attributed to the weak anti-localization effect (WAL). [8, 21, 29-33] According to the HLN theory[34]

$$\Delta\sigma = \sigma(B) - \sigma(0) = -\frac{\alpha e^2}{2\pi^2\hbar}\left[\ln\left(\frac{\hbar}{4Bel_\phi^2}\right) - \psi\left(\frac{1}{2} + \frac{\hbar}{4Bel_\phi^2}\right)\right]$$

where $\psi(x)$ is the digamma function and $l_\phi$ is the phase coherence length. $\alpha$ is a coefficient reflecting the strength of the spin-orbital coupling and magnetic scattering. For $Bi_2Te_3$ film, $\Delta\sigma$ at 1.8 K in the perpendicular field fits the HLN equation quite well in the low magnetic field and yields $\alpha = -0.2$ and $l_\phi = 709 nm$. Meanwhile, the temperature dependence of normalized resistance (R/R$_{min}$) for sample F has an upturn below 4 K with field 0 kOe and 20 kOe, as shown in Figure 4h, which is reminiscent of the electron-electron interaction in TI films. [8]

## 4. Conclusions

In summary, we have successfully demonstrated the MBE growth of Q$_2$Te$_3$ topological insulator films directly on GaAs (111) substrates. Similar to the results for Bi$_2$Se$_3$ grown on graphene/SiC(0001)reported by Liu et al., [25] spiral growth mode was observed for Q$_2$Te$_3$ films on Nv-GaAs (111)A substrate too. However, an step-bunched surface was observed for Q$_2$Te$_3$ films on V-GaAs (111)A substrates. Both XRD and HRTEM results indicate that the Nv-GaAs (111) substrate is better than a vicinal substrate to provide high quality Q$_2$Te$_3$ films. The tilted angle between Q$_2$Te$_3$ film and GaAs (111) substrate may be related to the misfit dislocation with a nonzero net out-of-plane Burgers vector component since Nagai's steps model is not applicable here due to the large lattice mismatch, [35, 36] and more interface investigations need to be done to establish the tilting mechanism in this kind of high-misfit heteroepitaxial systems. The observation of linear MR and WAL provides proofs for the possible TI behavior of as-grown Q$_2$Te$_3$ films. [10, 27] Hall effect measurements show that unintentional doping happens to as-grown Q$_2$Te$_3$ films: Sb$_2$Te$_3$ films always show p type conducting behavior, whereas, Bi$_2$Te$_3$ films always show n type conducting behavior, creating the potential to fabricate topological p-n junction on the same GaAs (111) substrate, which may offer a new platform to realize exciton condense in the interface. Further, our study paves a way to integrate TI-based devices with present semiconductor optoelectronic devices, generating brand-new multifunction devices.

## Acknowledgements

The authors gratefully acknowledge the NSF financial support through Grant No. DMR-0520550, the National Basic Research Program of China (Grant No. 2013CB934600), the National Natural Science Foundation of China (Grant No. 11174007 & No. 11222434), the Penn State MRSEC under NSF grant DMR-0820404. We are grateful to Moses H. W. Chan and Meenakshi Singh for the help in the transport measurement.

**Electronic Supplementary Material**: Figures 1S-5S give the information of RHEED, more AFM, Hall bar, and the transport properties of Q$_2$Te$_3$ films on GaAs (111)A substrates.

**Table 1.** Samples category of $Q_2Te_3$ on GaAs (111) substrate

| Sample. | Q$_2$Te$_3$ & Thickness | Buffer Layer & Thickness | Substrate[a] | Misalignment Angle | Conduction Type |
|---|---|---|---|---|---|
| A | Bi$_2$Te$_3$ & 200 nm | GaAs (111)A & 330 nm | N type V-GaAs (111)A→ ($\bar{2}$11) | 3° | n |
| B | Sb$_2$Te$_3$ & 400 nm | GaAs (111)A & 330 nm | N type V-GaAs (111)A→ ($\bar{2}$11) | 3° | p |
| C | Bi$_2$Te$_3$ & 30 nm | GaAs (111)A & 3 nm | N type V-GaAs (111)A→ ($\bar{2}$11) | 3° | |
| **D** | **Bi$_2$Te$_3$ & 30 nm** | **GaAs (111)A & 3 nm** | *Semi-insulating Nv-GaAs (111)A* | **±0.1°** | **n** |
| E | Bi$_2$Te$_3$ & 200 nm | GaAs (111)B & 330 nm | N type V-GaAs (111)B→ ($2\bar{1}\bar{1}$) | 3° | n |
| F | Sb$_2$Te$_3$ & 400 nm | GaAs (111)B & 330 nm | N type V-GaAs (111)B→ ($2\bar{1}\bar{1}$) | 3° | p |

[a]V-GaAs (111) substrate was made by AXT Inc., whereas, Nv-GaAs(111) was made by Wafer Technology Ltd.

Figures

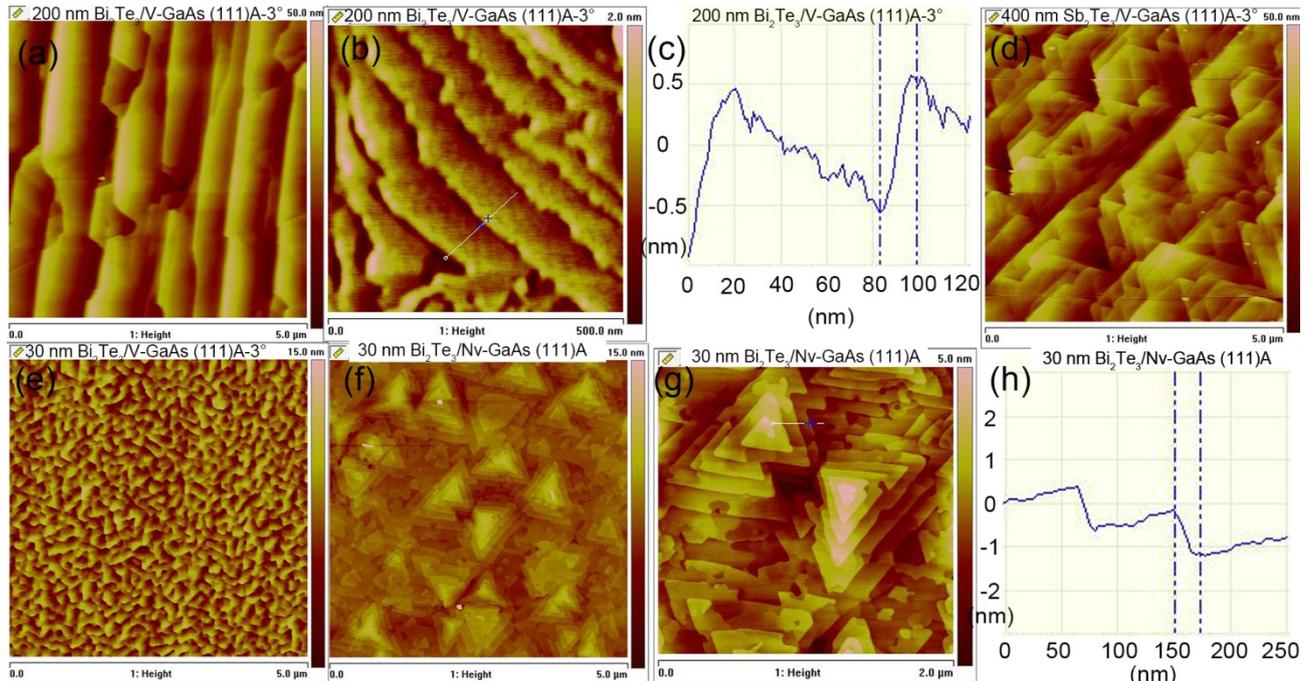

Figure 1. AFM images of surface morphology for as-grown Q$_2$Te$_3$ films on GaAs (111)A. a) Sample A in 5×5 µm$^2$, b) Sample A in 500×500 nm$^2$, c) the line profile corresponding to the line in b), d) Sample B in 5×5 µm$^2$, e) Sample C in 5×5 µm$^2$, f) Sample D in 5×5 µm$^2$, g) Sample D in 2×2 µm$^2$, h) the line profile corresponding to the line in g).

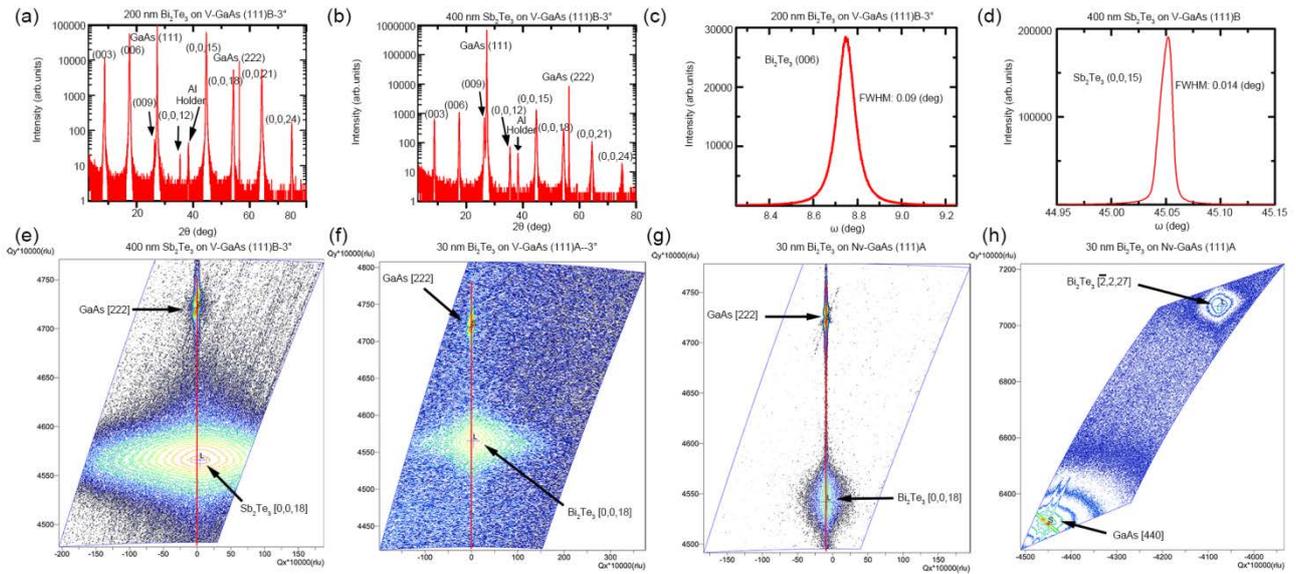

Figure 2. XRD patterns of as-grown $Q_2Te_3$ films. a) 2θ-ω scan of sample E, b) 2θ-ω scan of sample F, c) ω scan of sample E, d) ω scan of sample F, e) RSM mapping of sample F between GaAs [222] and $Sb_2Te_3$ [0,0,18], f) RSM mapping of sample C between GaAs [222] and $Bi_2Te_3$ [0,0,18], g) RSM mapping of sample D between GaAs [222] and $Bi_2Te_3$ [0,0,18], h) RSM mapping of sample D between GaAs [440] and $Bi_2Te_3$ [$\bar{2}2,27$] For RSM mapping, the Bartels monochromator was removed, hence Cu-Kα1 and Kα2 peaks are visible and a large Δλ streak is present through the substrate reflection.

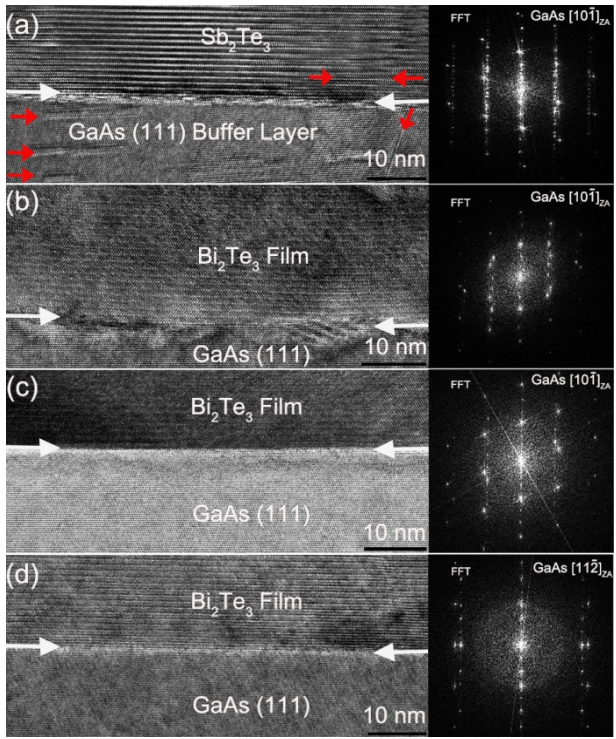

Figure 3. Cross-sectional HRTEM images of as-grown $Q_2Te_3$/GaAs (111) heterostructures. a) Sample F with the zone axis of GaAs [$10\bar{1}$], b) Sample C with the zone axis of GaAs [$10\bar{1}$], c) Sample D with the zone axis of GaAs [$10\bar{1}$], d) Sample D with the zone axis of GaAs [$11\bar{2}$]. The white arrows indicate the interface, the red arrows point defects, and the right rows show the corresponding FFT images of samples F, C and D; For all the FFT images, we see two sets of diffraction patterns with different symmetry: one is from GaAs substrate, and the other is from $Q_2Te_3$ film.

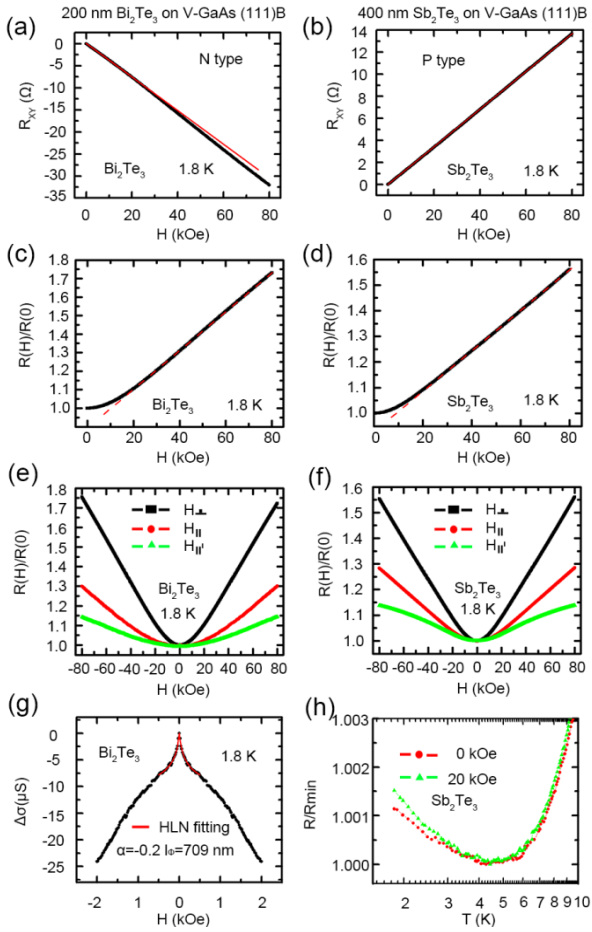

Figure 4. Transport properties of as-grown $Q_2Te_3$ films. a,b) Hall resistance versus magnetic field at 1.8 K for samples E and F, respectively, and the red solid line is the linear fitting, c,d) MR change in perpendicular magnetic field (black solid line) for samples E and F, respectively, and the red dashed line is the linear fitting one, e,f) MR change in both perpendicular and parallel magnetic fields configuration for samples E and F, g) Normalized magnetoconductance in perpendicular magnetic field between -2 and 2 kOe for sample E and the red solid line is the fitting curve with HLN equation at T=1.8K, h) Temperature dependence of normalized resistance($R/R_{min}$) at H = 0 kOe and 20 kOe for sample F, where $R_{min}$ is the minimum value of resistance. We use three different magnetic field configurations, $H_\perp$ denotes magnetic field perpendicular to the surface of the thin film, while $H_\parallel$ and $H_{\parallel'}$ denote an in-plane magnetic field perpendicular and parallel to the excitation current, respectively.

# Electronic Supplementary Material

# Molecular Beam Epitaxial Growth of Bi$_2$Te$_3$ and Sb$_2$Te$_3$ Topological Insulators on GaAs (111) Substrates: A Potential Route to Fabricate Topological Insulator p-n Junction

Zhaoquan Zeng[1,†], Timothy A. Morgan[1], Dongsheng Fan[1,2], Chen Li[1], Yusuke Hirono[1], Xian Hu[1], Yanfei Zhao[3], Joon Sue Lee[4], Jian Wang[3,4](✉), Zhiming M. Wang[1,5,6](✉), Shuiqing Yu[1,2], Michael E. Hawkridge[1], Mourad Benamara[1], and Gregory J. Salamo[1]

[1] Arkansas Institute for Nanoscale Material Sciences and Engineering, University of Arkansas, Fayetteville, AR 72701, USA
[2] Department of Electrical Engineering, University of Arkansas, Fayetteville, AR 72701, USA
[3] International Center for Quantum Materials, School of Physics, Peking University, Beijing, 100871, China
[4] The Center for Nanoscale Science and Department of Physics, The Pennsylvania State University, University Park, PA 16802, USA
[5] State Key Laboratory of Electronic Thin Films and Integrated Devices, University of Electronic Science and Technology of China, Chengdu 610054, China
[6] Engineering Research Center for Semiconductor Integrated Technology, Institute of Semiconductors, Chinese Academy of Science Beijing 100083, China
[†] Present address: Electrical and Computer Engineering Department, The Ohio State University, Columbus, Ohio 43210, USA

## 1. RHEED Evolution:

Figure 1S shows the typical RHEED evolution for Bi$_2$Te$_3$ films grown on GaAs (111)A substrates. Figure 1S(a-b) show the RHEED patterns of GaAs (111)A surface after deoxidizing and growth of GaAs buffer layer. After several minutes at deoxidizing temperature of 590 °C, the 2×2 reconstruction gradually appears, and it exists during the whole GaAs buffer layer growth. The clear 2×2 reconstruction and sharp streak indicate that a typical GaAs (111)A clean surface is formed[1]. While the substrate was cooled down to 250 °C, the 2×2 RHEED pattern disappeared, and 1×1 pattern appeared (figure 1S(c-d)). It can be attributed to the predeposition of Te, providing a Te rich growth condition for Bi$_2$Te$_3$. When the Bi shutter was open, the sharp streak became dim and shrank inside slowly (not shown here), which indicated the epitaxial growth of c-axis oriented Bi$_2$Te$_3$ film. As Bi$_2$Te$_3$ grew, the streaky RHEED patterns became sharp, indicating a flat surface. After epilayer growth for scores of nanometers, a set of sharp streaky pattern with clear Kikuchi lines was observed (figure 1S(e-f)), suggesting good crystallinity. It should be mentioned that the 1×1 surface persisted during the whole epilayer growth. Comparing the RHEED patterns of GaAs (111)A and Bi$_2$Te$_3$ epitaxy film, it can be concluded that relaxed Bi$_2$Te$_3$ (001) film has been grown on GaAs (111)A surface with an overlapped in-plane epitaxial relationship of Bi$_2$Te$_3$ $[10\bar{1}0]$//GaAs $[11\bar{2}]$ and Bi$_2$Te$_3$ $[11\bar{2}0]$//GaAs $[10\bar{1}]$.

Figure 2S shows the typical RHEED evolution for Bi$_2$Te$_3$ grown on GaAs (111)B substrates. Figure 2Sa shows the RHEED patterns of GaAs (111)B surface right after deoxidization at 590 °C, a typical $\sqrt{19}\times\sqrt{19}$ reconstructed surface of GaAs (111)B at high temperature[1] (the other direction is not shown here). During cooling down to 250 °C, a transition from $\sqrt{19}\times\sqrt{19}$ to 2×2 happened as shown figure 2Sb. Once Te shutter was open, the 1×1 surface appeared as figure 2Sc. After scores of nanometers epitaxial growth of Bi$_2$Te$_3$ film, a set of RHEED pattern with Kikuchi lines appeared as figure 2Sd, indicating good crystal quality.

---

Address correspondence to Jian Wang, email: jianwangphysics@pku.edu.cn ; Zhiming M. Wang, email:zhmwang@semi.ac.cn

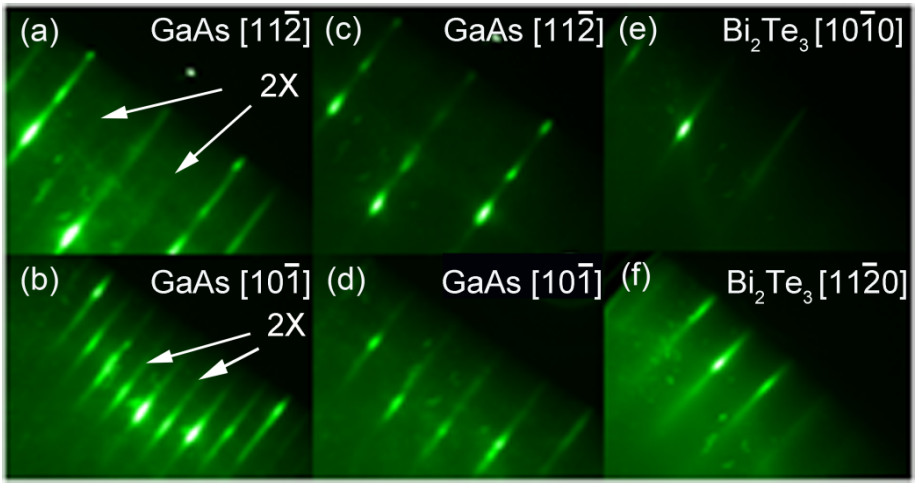

Figure 1S. RHEED patterns taken from a,b) GaAs (111)A-(2×2) after deoxidizing and the growth of buffer layer, c,d) GaAs (111)A after Te predeposition, and e,f) Bi₂Te₃ epilayer after scores of nanometers growth.

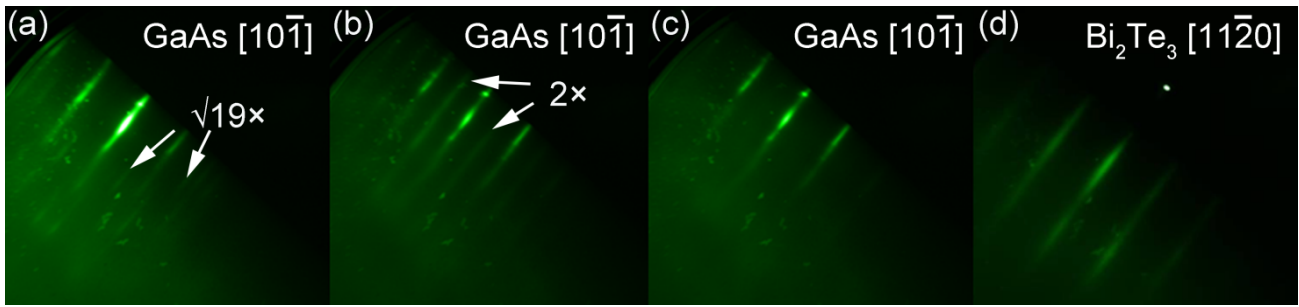

Figure 2S. RHEED patterns taken from a) GaAs (111)B-(√19×√19) right after the growth of buffer layer at 590 °C, b) GaAs (111)B (2×2) after cooling down to 250 °C, c) GaAs (111)B (1×1) after Te predeposition, and d) Bi₂Te₃ epilayer after scores of nanometers epitaxial growth.

**Table 1S. Samples category of $Q_2Te_3$ on GaAs (111)A substrate**

| Sample | $Q_2Te_3$ & Thickness | Buffer Layer & Thickness | Substrate[a] | Misalignment Angle | Conduction Type |
|---|---|---|---|---|---|
| G | $Bi_2Te_3$ & 100 nm | GaAs (111)A & 3 nm | Semi-insulating Nv-GaAs (111)A | ±0.1° | n |
| H | $Bi_2Te_3$ & 100 nm | GaAs (111)A & 3 nm | Semi-insulating V-GaAs (111)A → ($\bar{2}$11) | 3°±0.1° | |
| I | $Bi_2Te_3$ & 100 nm | GaAs (111)A & 3 nm | Semi-insulating V-GaAs (111)A → ($\bar{1}\bar{1}$2) | 2°±0.1° | |
| J | $Bi_2Te_3$ & 100 nm | GaAs (111)A & 3 nm | Semi-insulating V-GaAs (111)A → ($\bar{1}\bar{1}$2) | 1°±0.1° | |

[a] Here, both V-GaAs (111)A and Nv-GaAs (111)A substrates were made by Wafer Technology Ltd.

## 2. More AFM Investigations:

In order to further clarify the effect of the GaAs (111) surface on the growth mode of as-grown $Q_2Te_3$ films, more samples were grown, as shown in Table 1S. Figure 3S shows additional AFM results on the morphology of thicker $Q_2Te_3$ films grown on both V-GaAs (111)A and Nv-GaAs(111)A substrates. Figures 3S(a-c) are the AFM images of sample G, and it is very clear that the QL steps with a height of around 1 nm were formed over the whole surface, indicating that spiral growth mode can be maintained during the epitaxial growth of $Q_2Te_3$ film. Figure 3S(d-f) are the AFM images of sample H, and big terraces and steps were formed on the surface due to the step-flow growth mode. Even for samples I and J, with smaller miscut angle (2° and 1°), the step-bunched surface morphology was still kept, as shown in figures 3S(g-l).

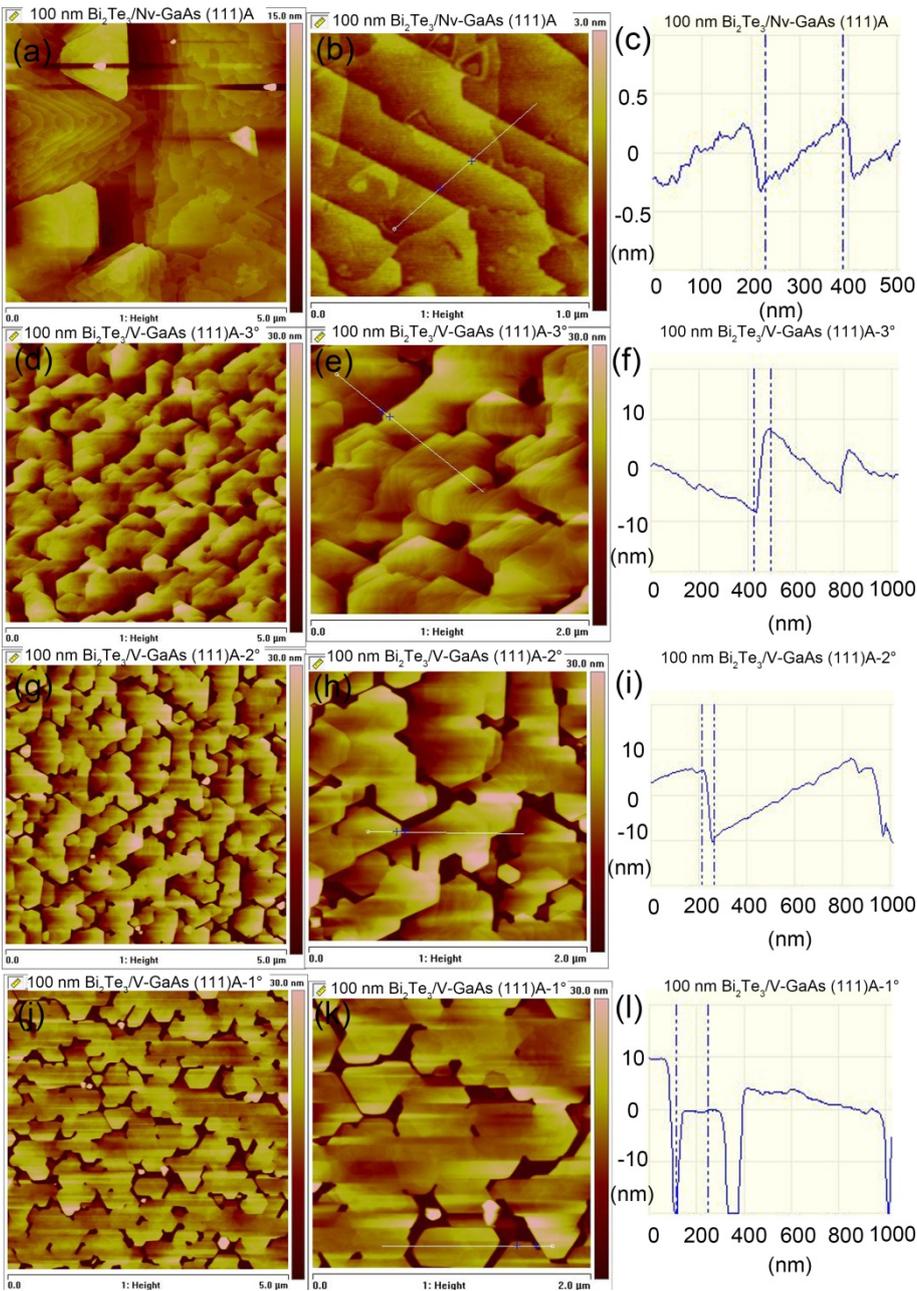

Figure 3S. AFM images of surface morphology for as-grown $Q_2Te_3$ films. a) Sample G in 5×5 μm², b) Sample G in 500 nm×500 nm, c) the line profile corresponding to the line in b), d) Sample H in 5×5 μm², e) Sample H in 2×2 μm², f) the line profile corresponding to the line in e), g) Sample I in 5×5 μm², h) Sample I in 2×2 μm², i) the line profile corresponding to the line of h), j) Sample J in 5×5 μm², k) Sample J in 2×2 μm², l) the line profile corresponding to the line in k).

### 3. Hall Bar

All the transport measurements are based on the same Hall bar structure as shown in Figure 4S. Figure 4Sa is the schematic diagram (not proportional), and Figure 4Sb is the corresponding optical microscopy image. The standard photo-lithography method was used for fabricating the Hall bar. [2] Positive photoresist S1813 was spun at 4000 rpm for 45 seconds on $Q_2Te_3$ films, followed by 110 ℃ baking for 60 seconds. With a mask of Hall bar pattern, the photoresist-coated sample was exposed to ultraviolet light (365 nm wavelength) with exposure power of 8 mW/cm² for 7 seconds. The exposed part of the photoresist was removed after 40 seconds of developing (MF CD-26 developer). Then, bare area of $Bi_2Te_3$ film with no photoresist was wet-etched with 1 g of potassium dichromate in 10 ml of sulfuric acid and 520 ml of deionized (DI) water, whereas, diluted nitric acid ($HNO_3$: $H_2O$ = 1:1) was used for $Sb_2Te_3$ film. The expected etching rate for both $Bi_2Te_3$ and $Sb_2Te_3$ films is about 10 nm per minute, and DI water rinsing is needed.

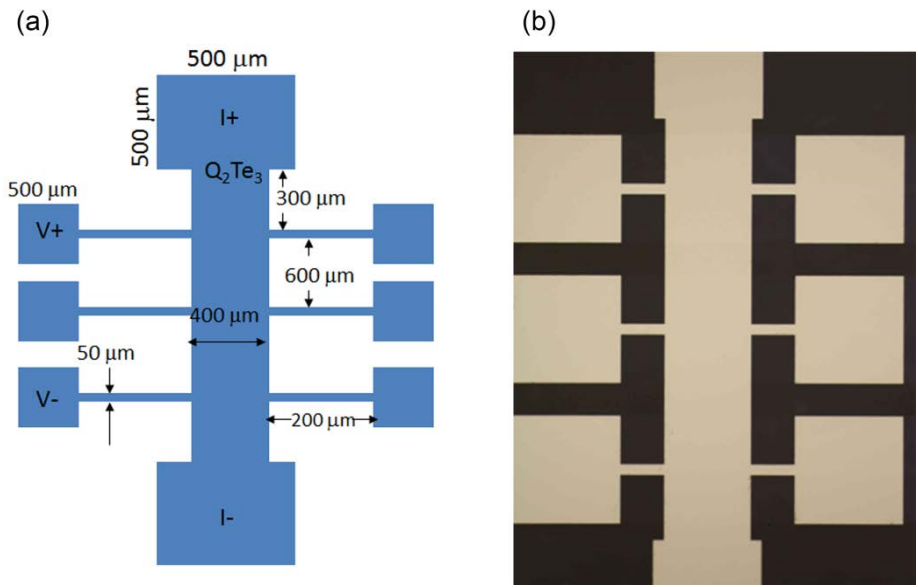

Figure 4S. Hall bar schematic diagram (not proportional) a) and the corresponding optical microscopy image b).

### 4. Transport Properties of Q2Te3 films on GaAs (111)A Substrates

Figure 5S shows the typical transport data for $Q_2Te_3$ films on GaAs (111)A substrates. Low temperature Hall measurements indicates that samples B and D show p and n type conductivity, respectively, which support the conclusion that all the as-grown $Sb_2Te_3$ films show p type conducting behavior and all the as-grown $Bi_2Te_3$ films show n type conducting behavior. Similar to the case of $Q_2Te_3$ films on GaAs (111)B, nonlinear Hall behavior was also observed over 20 kOe for both samples B and D at 1.8 K, as shown in figures 5S(a-b), which is likely an indication of more than one transport channel (surface and bulk). Based on the linear behavior below 20 kOe, the carrier density can be estimated as $3.81×10^{19}$ cm$^{-3}$ and $2.04×10^{19}$ cm$^{-3}$ for samples B and D, respectively. Compared to the result in Figure 4h, a stronger upturn was observed for sample B with fields 20 and 80 kOe, as shown in Figure 5Sc. This is also observed for sample D below 10 K as show in figure 5Sd. Figures 5Sc and 5Sd show the normalized resistance upturn, which is enhanced by applying perpendicular magnetic field. The behavior is reminiscent of the electron-electron interaction in TI films.[3] Additionally, the WAL induced magnetoconductance peak at small field was also observed for both sample B and D, as shown in Figures 5S(e-f).[3-7] It should be noted that the magnetoconductance peak was well fitted by HLN function.[3] For $Sb_2Te_3$ the fitting yields $\alpha = -1$ and $l_\phi = 101 nm$ and for $Bi_2Te_3$ $\alpha = -0.18$ and $l_\phi = 225 nm$ as shown in Figure 5S e-f respectively, suggesting its quantum origin [3]. The MR for three orientations of magnetic field was also measured at 1.8 K for both sample B and D, and it is obvious that the MR for perpendicular field is much higher than those for parallel fields, as shown in Figures 5Sg-h. A near parabolic MR was observed for sample B in perpendicular magnetic field, and it is suppressed when the field is parallel to the current. However, a linear MR was observed for sample D with perpendicular field above 30 kOe, which may be attributed to the QLMR of the surface states [8].

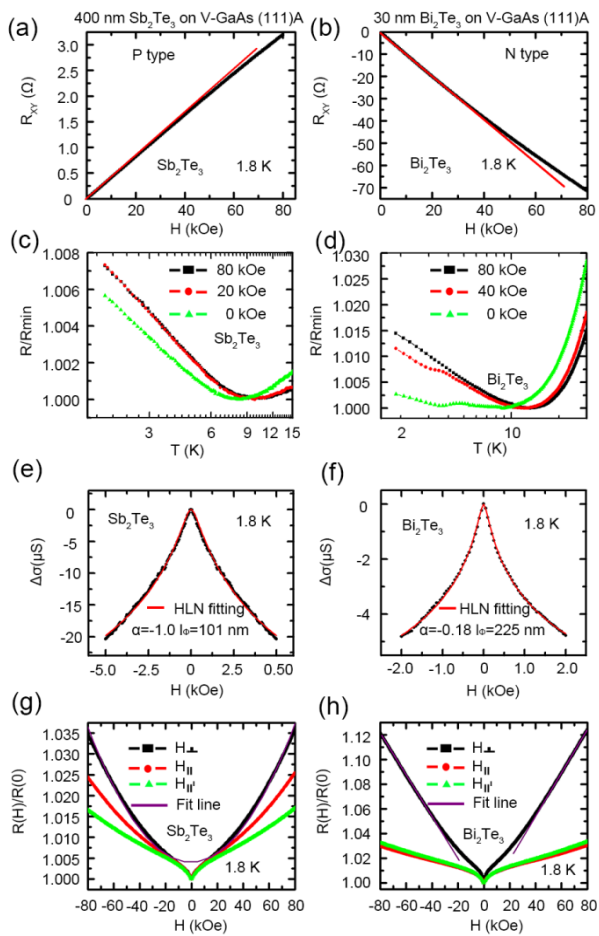

Figure 5S. Transport properties of as-grown $Q_2Te_3$ films on GaAs (111)A substrates. a,b) Hall resistance versus magnetic field (black solid line) at 1.8 K for samples B and D, respectively, and the red solid line is the linear fit, c) Temperature dependence of normalized resistance($R/R_{min}$) at H = 0 kOe, 20 kOe and 80 kOe for sample B, where $R_{min}$ is the minimum value of resistance, d) ) Temperature dependence of normalized resistance($R/R_{min}$) at H = 0 kOe, 4 kOe and 8 kOe for sample D, e) Normalized magnetoconductance in perpendicular magnetic field at T=1.8K for sample B. The red solid line is the fitting curve with HLN equation, f) Normalized magnetoconductance in perpendicular magnetic field at T=1.8K for sample D. The red solid line is the fitting curve with HLN equation, g,h) MR change in both perpendicular and parallel magnetic fields configuration for samples B and D at 1.8 K, and the violet line is the linear fit. We use three different

magnetic field configurations, $H_\perp$ denotes magnetic field perpendicular to the surface of the thin film, while $H_\parallel$ and $H_{\parallel'}$ denote an in-plane magnetic field perpendicular and parallel to the excitation current, respectively.